\documentclass[12pt]{article}

\usepackage[latin1]{inputenc}
\usepackage{graphicx}
\usepackage[T1]{fontenc}

\usepackage{amsmath}
\usepackage{amssymb} 
\usepackage{amsfonts}
\usepackage{mathtools} 
\usepackage{bm}
\usepackage{mathrsfs}
\DeclareMathOperator{\erfc}{erfc}
\newcommand{\diff}{\,\mathrm{d}}

\newcommand{\DE}{\Delta E}

\title{Chopping time of the FPU $\alpha$-model}

\author{ A. Carati\footnote{ Universit\`a di Milano, Dipartimento di
    Matematica, Via Saldini 50,  20133 Milano (Italy).   
E-mail: {\tt andrea.carati@unimi.it}  } 
  \and
  A. Ponno\footnote{Universit\`a degli Studi di Padova,
    Dipartimento di Matematica ``Tullio Levi-Civita'', 
    Via Trieste 63, 35121 Padova
    (Italy). E-mail: {\tt ponno@math.unipd.it }}
}

\date{\today}

\begin{document}

\maketitle

%\vskip 1 truemm
%\centerline{ Universit\`a di Milano, Dipartimento di Matematica }
%\vskip 1 truemm
%\centerline{Via Saldini 50,  20133 Milano (Italy)}
%\vskip 1 truemm
%\centerline{ E-mail: {\tt carati@mat.unimi.it} }
%\vskip 1 truecm

\begin{abstract}
We study, both numerically and analytically, the time needed to
observe the breaking of an FPU $\alpha$-chain in two 
or more pieces, starting from an unbroken configuration at a given temperature.
It is found that such a ``chopping'' time is given by a formula that,
at low temperatures, is 
of the Arrhenius-Kramers form, so that the chain does not break up on
an observable time-scale. The result explains why the study of the FPU
problem is meaningful also in the ill-posed case of the $\alpha$-model.  
\end{abstract} 
\noindent  
\vskip 1truecm
%\vfill  %%% DA METTERE IN FONDO 
%\noindent PACS numbers: 03.50.D, 03.65.B,  02.30.M 
%\par

%%% DA CANCELLARE
%\noindent
%Running title: Chopping time for the FPU $\alpha$-model  
%\vskip 3.truecm
%\noindent
%\eject %%%% Da cancellare se non si vuole l'abstract in una pagina

\section{Introduction} 

As is well known, the so-called Fermi-Pasta-Ulam (FPU) model was introduced in 1954 \cite{FPU}, and consists in a
one-dimensional chain of $N$ identical particles, interacting through
a nearest neighbor, non quadratic potential $V(x)$, $x$ denoting the
inter-particle distance. The problem FPU were interested in was the
characterization of the relaxation path of the system to its
micro-canonical equilibrium. Such an issue, known as the FPU problem, is related to, but is not the main focus of our work, and the interested reader is referred, for example, to the recent works \cite{BLP}--\cite{GGPP}.

In the present paper we deal with the problem of determining the
life-time of a chain with pairwise potential $V(x)$ that is not lower
bounded. In particular, we study the so-called FPU $\alpha$-model,
i.e. the model with cubic potential
\begin{equation}
\label{eq:V}
V(x) = \frac 12 x^2 + \frac \alpha3 x^3\ \ (\alpha\neq0)\ ,
\end{equation}
which was among those originally considered by FPU \cite{FPU} and widespread in the literature of the field since then. As is well known, potentials which are not bounded from below, pose
problems from the point of view of both statistical mechanics and
dynamics. Indeed, due to the lack of compactness of the
constant energy surface, the micro-canonical (as well as the
canonical) measure and the global (in time) solution of the equations
of motion do not exist, with two unpleasant consequences:
\begin{itemize}
\item[i)] the equilibrium thermodynamics of the system is not defined;
\item[ii)] the system is expected to (and actually does) blow-up in a finite
time.
\end{itemize}
In fact, the cubic potential (\ref{eq:V}) displays a potential well of
finite height $\DE=V(-1/\alpha)=(6\alpha^2)^{-1}$ on the left/right of $x=0$, and tends to $-\infty$ as
$x\to\mp\infty$, according to whether the sign of $\alpha$ is
$\pm1$. This implies that, for total energy values above the threshold
$\DE$, at least one particle can escape to infinity, causing a local
breakdown of the chain.  Thus, if one works at fixed specific energy
(energy per particle), for a sufficiently large number $N$ of particles
the chain breaks down into pieces of finite length, regardless of the specific energy value.
Equivalently, if $N$ is very large and one extracts the particle velocities from a Maxwell
distribution at temperature $T=1/\beta$, the kinetic energy of
some particles will be larger than $\Delta E$, so that, no matter how small
$T$ is, the chain should end up ``chopped''. On the other hand, there are many numerical
studies in the literature, starting from that of FPU, devoted to this problem, where the chain is
not observed to break down, and is even observed to reach and persists
in a state characterized by quasi equipartition of the energy,
at least when one starts with unbroken configurations of the chain and the temperature is small enough.

More in general, we stress that a chain would end up broken with any realistic, short range, interaction potential tending to a constant value out of a well of a finite depth and width, such as the Lennard-Jones and the Morse ones. That is why we think that the study of
the \emph{chopping time} (i.e. the time needed to the chain to break up into two or more pieces) is of physical relevance. Just to make a nontrivial example, think of the DNA
dynamics \cite{Pe04,Pe06} where, if the temperature is low enough, the double helix
does not unbind, notwithstanding it should do that by heuristic
arguments similar to those reported above. This means that the system remains trapped in
an unlikely state on a long term, which also displays analogies with
the behavior of glasses; some comments about this point are deferred
to the last Section.

However, for the study of the chopping time, the choice of the
unphysical $\alpha$-FPU model is not ``odd'', but, on the contrary,
benefits of certain advantages in numerical simulations, for example a
clear cut determination of the blow-up, signaled by an overflow, as
shown below. We observe moreover that the potential (\ref{eq:V}) is
the third order expansion of any realistic potential around the
minimum point of its attracting well. In this sense, the phenomenology
of any chain chopping should be well described by the simple
$\alpha$-model.

The results we found are the following. By suitably choosing random initial conditions (to be specified below in Section 2), we estimate numerically
the mean $t_c$ and the standard deviation $\delta t_c$ of the chopping time as a function of the inverse
temperature $\beta$ of the system, for two different (large) values of the number $N$ of particles. Moreover, a theoretical estimate of the probability distribution of the chopping time is also obtained, which leads to the law
\begin{equation}
 \label{eq:1}
    t_c =\frac{A}{1-(1-\mathsf{p})^N}=\frac
    {A}{1-e^{N\ln(1-\mathsf{p})}}\ ,
\end{equation}
where $A$ is the free parameter of the theory, whereas $\mathsf{p}$ can be expressed in terms of the complementary
error function\footnote{We recall that the complementary error
  function is defined as
\[
 \erfc(x)=(2/\sqrt{\pi})\int_x^{+\infty}e^{-t^2}\diff t \ .
\]} 
$\erfc(x)$ as follows
\begin{equation}
\label{eq:2}
\mathsf{p}=\erfc\left(\sqrt{\beta\Delta E}\right)\ .
\end{equation}
Here $\beta$ is the inverse temperature of the system (determined by the initial conditions) and $\DE$
is the height of the potential well.
The quantity $\mathsf{p}$, defined in (\ref{eq:2}), will be shown to be the probability that
a local blow-up takes place somewhere in the chain. The theoretical law
(\ref{eq:1}) is plotted against the numerical data in Figure~\ref{fig:1} below.

Notice that $t_c\to A$ both in the high temperature limit $\beta\Delta
E\to0$ (which implies $\mathsf{p}\to1$), and $N\to+\infty$ at any
fixed value of $\beta\DE$, so that the parameter $A$ has the meaning of expected chopping time in the thermodynamic and/or high temperature limit.  On the other hand, if $\beta\DE$ is so
large that $N\mathsf{p}\ll1$, one gets the asymptotic behavior
\begin{equation}
\label{eq:3}
t_c\sim\frac{A}{N\mathsf{p}}\sim\frac{A}{N}\ \sqrt{\pi\beta\Delta
  E}\ e^{\beta\Delta E}\ ,
\end{equation}
i.e. a law of the Arrhenius-Kramers type. In other terms the chopping time
increases exponentially fast with the inverse temperature $\beta$.
This explains why any finite chain (also of a macroscopic size) may
remain frozen in an unbroken state, for a sufficiently low (but
finite) temperature.

According to the theory, the predicted standard deviation $\delta t_c$ on the chopping time is
\begin{equation}
\label{eq:stdev}
\delta t_c=A\ 
\frac{e^{N\ln\sqrt{1-\mathsf{p}}}}{1-e^{N\ln(1-\mathsf{p})}}\ .
\end{equation}
Thus, when the chain chopping becomes a rare event ($N\mathsf{p}\ll1$), the expected
chopping time $t_c$ undergoes very large fluctuations, with 
$\delta t_c\sim t_c$; on the other hand $\delta t_c\to0$ very quickly in the
limit $N\to\infty$ or $\beta\DE\to0$. The theoretical law
(\ref{eq:stdev}) is plotted against the numerical data in Figure~\ref{fig:2} below.

The paper is organized as follows.  In Section~\ref{sec:3} the results
obtained by numerical integrating the equation of motions for a system
of $N=65,535$ and one of $N=1,048,575$ particles are reported, and the
agreement with the law (\ref{eq:1}) for the chopping time is
illustrated. Such a law is deduced in Section~\ref{sec:2} by
considering the dynamics of the chain at discrete time intervals and
defining the probability $P$ of chain chopping during each time
step. A few plausible hypotheses on such a probability allow then to
simply compute the expected chopping time. The constant $A$ appearing
in (\ref{eq:1}) cannot be deduced from the analytic computations, but
it is nevertheless estimated by the numerical results of
Section~\ref{sec:3}. Further comments are deferred to
Section~\ref{sec:4}.
\begin{figure}[!t]
  \begin{center}
    \includegraphics[width=\textwidth]{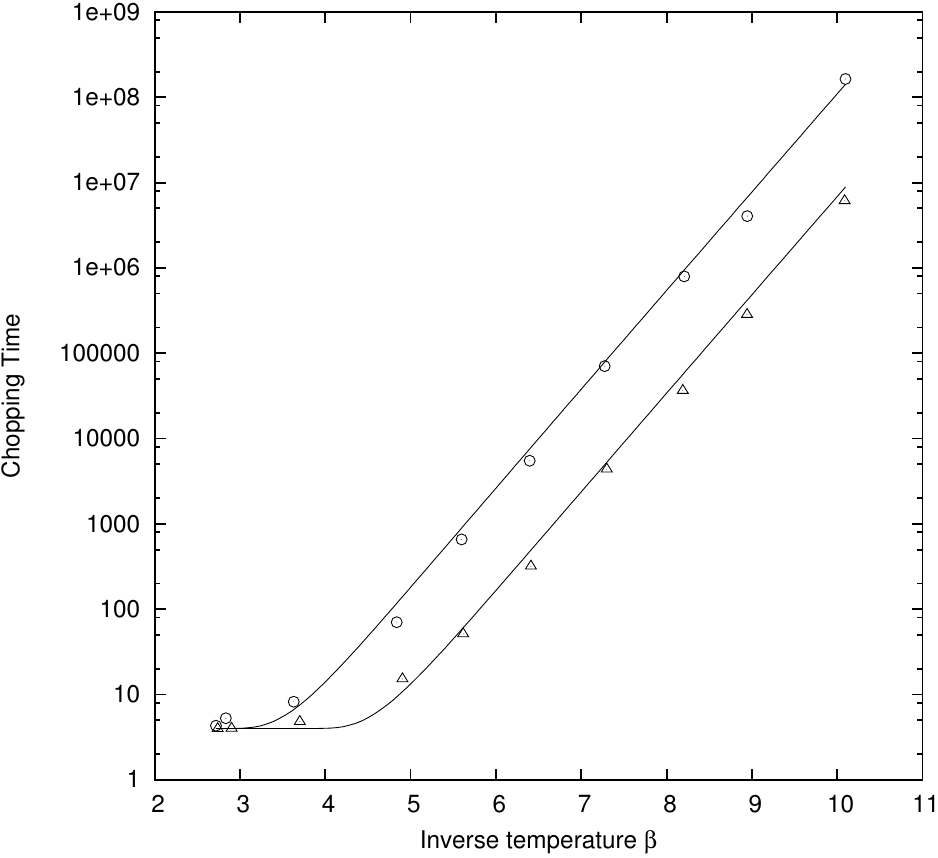}
  \end{center}
  \caption{\label{fig:1} The chopping time $t_c$, as a
    function of $\beta$, in semi logarithmic scale. Circles refer
    to $N=65,535$ particles, while triangles refer to $N=1,048,575$. 
    Solid lines are the plots of the analytic formula (\ref{eq:1}) with
    $\mathsf{p}=\erfc\left(\sqrt{\beta\Delta E}\right)$, $A=4.0$
    and $\Delta E=2.57$.  }
\end{figure}

\begin{figure}[!t]
  \begin{center}
    \includegraphics[width=\textwidth]{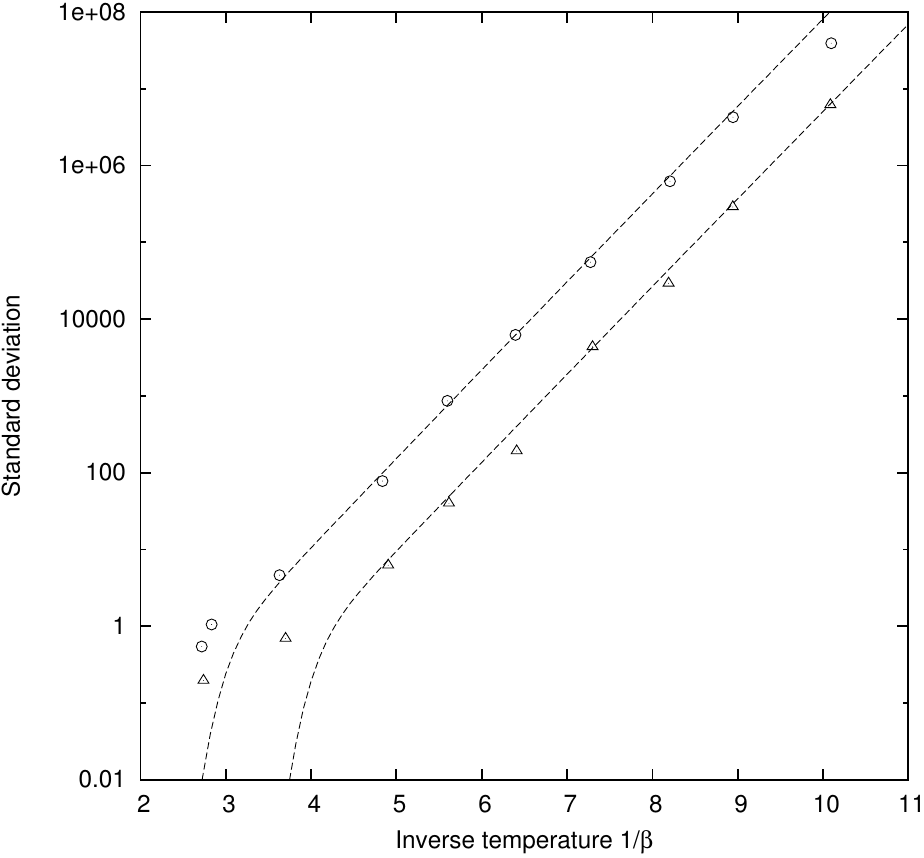}
  \end{center}
  \caption{\label{fig:2} The standard deviation $\delta t_c$, as a
    function of $\beta$, in semi logarithmic scale. Circles refer
    to $N=65,535$ particles, while triangles refer to $N=1,048,575$. 
    Dotted lines are the plots of the analytic formula
    (\ref{eq:stdev}), with
    $\mathsf{p}=\erfc\left(\sqrt{\beta\Delta E}\right)$, $A=4.0$
    and $\Delta E=2.57$.  }
\end{figure}
\section{Numerical computations}\label{sec:3} 

To the purpose of numerical integration we have set
$\alpha=0.25(=1/4)$ in the potential (\ref{eq:V}). Notice that such a
value is by no means special because, by a suitable rescaling of the
variables, one can always reduce to this case. Thus, the
Hamiltonian studied is 
\[ 
H = \frac 1{2} \sum_{j=1}^N p_j^2 + \frac 12 \sum_{j=0}^N (q_{j+1}
- q_j)^2 + \frac 1{12} \sum_{j=0}^N (q_{j+1} - q_j )^3 \ ,
\] 
with fixed ends: $q_0=q_{N+1}=0$.  The numerical integration of the
equations of motion was performed by using the standard Verlet
algorithm \cite{Ver67}, with a time--step equal to $0.025$, which ensures the
energy conservation up to a relative error less than $10^{-3}$ in all
the computations performed.

We considered two different numbers $N$ of particles, namely
$N=65,535$ and $N=1,048,575$ respectively, and specific energy values
in the range $0.1\div1$. In dealing with such large numbers of
particles, care has to be taken in choosing the initial data,
otherwise, in the chosen energy range, the chain is found to be almost
immediately chopped. We remark that the chopping phenomenon takes
place when the relative distance between a 
couple of neighboring particles is so large (namely larger than
$|-1/\alpha|=4$) that the
force between them is repulsive and the two halves of the chain
separate out. For such a reason we decided to set all the initial
positions $q_j^0=0$, for all $j=1,\ldots, N$, while the initial
momenta $p_j^0$ are extracted from a Maxwell distribution at inverse
temperature $\beta$, taking care to reject all those values with a
kinetic energy too large. More precisely, if it happens that $\Big( 
p_j^0 \Big)^2 > \Delta E$, we extract $p_j^0$ again and again until
such a condition no longer holds. In this way, we are sure to start in the
(unlikely) state in which the chain is unbroken. Then we begin the
numerical simulation, going on with it up to the time $\tilde t$ such that
\begin{equation}
\label{eq:tilt}
q_{j+1}(\tilde t\,) - q_j(\tilde t\,) < -6
\end{equation}
for at least one $j$. The threshold value $-6$ for the inter-particle
distance is conventionally fixed among the possible ones smaller than
the local maximum point $-1/\alpha=-4$. 
As we have just
remarked, when this condition holds, the force between the two
particles $q_{j+1}$ and $q_j$ becomes repulsive and the two pieces of
the chain move apart from each other.
For times longer than $\tilde t$ one always gets an overflow error
very quickly, because, in 
our model, the repulsive force grows very fast as the distance
increases. This means that there exists a local blow-up
time $t_b\in\mathbb{R}$ such
that
\[
 \lim_{t\to t_b^-} \sup_j |q_{j+1}-q_j| = +\infty \ .
\]
The local blow-up time time $t_b$ would be the actual chopping time of
the chain. However, as just explained above, to all practical
purposes,  
$t_b$ is well approximated by the time $\tilde t$ defined by the relation
(\ref{eq:tilt}), with the obvious advantage of avoiding to stop the
computation at the overflow, which in turn allows to implement
cyclical runs.  
%\begin{figure}[!t]
%  \begin{center}
%    \includegraphics[width=\textwidth]{fig_chop2}
%  \end{center}
%  \caption{\label{fig:2} The chopping time standard deviation $\delta
%    t_c$, as a function of $\beta$, in semi-logarithmic
%    scale. Triangles refer to $N=65,535$, while circles refer to
%    $N=1,048,575$. The fluctuations are comparable with the average;
%    see Figure~\ref{fig:1}. }
%\end{figure}
Indeed, since the time $\tilde t$ (as well as $t_b$) depends on the
initial data, it is 
a random variable varying from one numerical experiment to the
other. As a consequence, we define the \emph{expected chopping time
  $t_c$} as the average of $\tilde t$ with respect to the initial
data. In practice, once fixed 
the value of the inverse temperature $\beta$, one repeats the
numerical experiment with a set 
of $M$ different initial conditions. Then one gets a sample of
different values $\tilde t_k$, $k=1,\ldots,M$, and estimates $\tilde t$ by
the its (sample) average
\[
t_c = \frac 1{M} \sum_{k=1}^M \tilde t_k \ .
\] 
In our simulations we use $M=25$, which is the largest number we
were able to reach with the computational power available.

The result are shown in Figure~\ref{fig:1}, where we report in
semi-logarithmic scale the expected chopping time $t_c$ as a function of
$\beta$. The exponential behavior becomes evident for
values of $\beta$ larger than about 6. The solid lines are the plots of the
formula (\ref{eq:1}), with $\mathsf{p}=\erfc(\sqrt{\beta\Delta E})$, 
for the two mentioned values of $N$ and an empirical value $\Delta E=
2.57$, which 
agrees well with the theoretical one, namely $\Delta
E=(6\alpha^2)^{-1}=2.66$. The value of the constant $A$ (a free
parameter in the theory) turns out to be equal to $4$. 

In the same way, we compute the chopping time (sample) standard 
deviation $\delta t_c$ defined by
\[
\delta t_c = \sqrt{\frac 1{M} \sum_{k=1}^M (\tilde t_k - t_c)^2 } .
\] 
The result, shown in Figure~\ref{fig:2}, is particularly
relevant. The standard deviation is of the same order of magnitude of the
average when the chopping of the chain is a \emph{rare event}. In
other terms, the fluctuations become very large by decreasing the
temperature, as predicted by the formula (\ref{eq:stdev}). However,
the latter formula does not fit the numerical data at high
temperatures. This is possibly due to different reasons; in any case the
smallness of the sample prevents to compute small values of standard
deviation with accuracy.

We finally stress that the small discrepancy between the optimal
numerical value of 
$\DE$ and the theoretical one, amounting to about $3\%$, might partly
depend on the fact that $\tilde t<t_b$ for any choice of the threshold 
value for the inter-particle distance, which we have fixed to $-6$. As
a consequence, the chopping time $t_c$ determined numerically turns
out to be an underestimate of the true one (the nasty sample average
of the overflow times). Another possible cause of such a discrepancy
will be discussed at the end of the next Section.

\section{Analytic estimate of $t_c$}\label{sec:2} 

Our analytic deduction of formula (\ref{eq:1}) is based
on the following assumptions and reasonings.  Let us consider the FPU
$\alpha$-chain with some measure on the initial conditions evolved on
consecutive time intervals of length $A$.  Let us call $P_k$ the
probability of the event $E_k$: \emph{chain chopping occurs in the
  time interval $\Delta t_k= [(k-1)A,kA]$}.  As anticipated above, by
chain chopping we mean the occurrence of at least one local blow-up,
namely the divergence, in a finite time, of at least one of the
inter-particle distances $r_n:=q_{n+1}-q_n$ to $-\infty$ (obviously,
the probabilistic nature of the event $E_k$ is inherited by the
evolution in time of the measure chosen on the initial data). One has
$P_k=1-Q_k$, where $Q_k$ denotes the probability of the complementary
event $\complement E_k$: \emph{no chain chopping on the time interval
  $\Delta t_k$ occurs}. The first simplifying hypothesis is introduced
here, assuming that the local blow-up events giving rise to the chain
chopping are mutually independent and occur with the same probability
for each particle pair.  In such a way one can write
$Q_k=(\mathsf{q}_k)^N$, $\mathsf{q}_k$ denoting the probability that
$r_n$ is lower bounded on the time interval $\Delta t_k$, for any
$n=1,\dots,N$. Such a hypothesis of statistical independence is more
plausible for free ends or periodic boundary conditions, while for
fixed ends (which is the case numerically considered here) some
boundary effect is expected, with a vanishing contribution as $N$ gets
larger and larger.  Now, writing $\mathsf{q}_k=1-\mathsf{p}_k$, one
gets
\begin{equation}
\label{eq:Pk}
P_k=1-(1-\mathsf{p}_k)^N=1-e^{N\ln(1-\mathsf{p}_k)}\ ,
\end{equation}
where $\mathsf{p}_k$ denotes the probability that $r_n\to-\infty$ on
the time interval $\Delta t_k$, for \emph{some} $n=1,\dots,N$. A second
fundamental hypothesis is here introduced, assuming that the
probability $\mathsf{p}_k$ is independent of $k$, i.e. of the specific
time interval $\Delta t_k$, depending instead only on its length
$|\Delta t_k|=A$. This is clearly a Markov-like hypothesis, equivalent
to assume that, up to the occurrence of the first local blow-up, the
measure on the initial data is just slightly modified by the flow, so
that the probability of local blow-up does not significantly change in
the course of time. Thus $\mathsf{p}_k\simeq\mathsf{p}_1$ for any $k$,
$\mathsf{p}_1$ being the probability that a local blow-up occurs
(i.e. $r_n\to-\infty$) on the time interval $[0,A]$, i.e. the
probability, according to the measure on the initial data, that the
local blow-up time is less than $A$. We shall see below how to compute
$\mathsf{p}_1(A)$. The consequence of this second hypothesis is that
the probability of chain chopping (\ref{eq:Pk}) on any time interval
$\Delta t_k$ is given by
\begin{equation}
\label{eq:PkP1}
P_k\simeq P_1=1-e^{N\ln(1-\mathsf{p}_1)}\ .
\end{equation}
The latter simplification allows us to get a simple expression for the
probability $\pi_n$ that the chain breaks down for the first time in
the $n$-th time interval $\Delta t_n=[(n-1)A,nA]$ and not before,
namely the geometric distribution \cite{Ross}
\begin{equation}
\label{eq:pin}
\pi_n=(1-P_1)^{n-1}P_1\ .
\end{equation}
The (mean) chopping time $t_c$ is then naturally defined as the
expected time one has to wait for the occurrence of the first local
blow-up of the chain, namely
\begin{equation}
\label{eq:tcdef}
t_c:=A\langle n\rangle:=A\sum_{n\geq1}n\pi_n\ .
\end{equation}
An elementary computation, making use of the properties of the
geometric series, yields
\begin{equation}
\label{eq:tcform}
t_c=\frac{A}{P_1}=\frac{A}{1-e^{N\ln(1-\mathsf{p}_1)}}\ ,
\end{equation} 
which has the form (\ref{eq:1}). We stress here that the latter
formula for the chopping time admits the two following limit
expressions. The first one holds if $\mathsf{p}_1$ is kept fixed and
$N\to+\infty$, i.e. in the ideal, or mathematical thermodynamic
limit. In this case, since $\ln(1-\mathsf{p}_1)<0$ one gets $P_1\to1$
and $t_c\to A$. Such a result is independent of any detail of the
system and actually defines the up to now arbitrarily chosen time
unit: \emph{A is the chopping time of the infinite chain}. This is a
quantity we are not able to compute analytically and is thus left as a
free parameter of the theory, to be numerically determined. The other
interesting limit of formula (\ref{eq:tcform}) is obtained when $N$ is
thought of as fixed, though as large as needed, while $\mathsf{p}_1$
is so small that $N\mathsf{p}_1\ll1$ (which is possible because
$\mathsf{p}_1$ depends on the temperature). In this case one has
$P_1\simeq 1-e^{-N\mathsf{p}_1}\simeq N\mathsf{p}_1$, and $t_c\simeq
A/(N\mathsf{p}_1)$. The latter asymptotic expression is clearly of the
form (\ref{eq:3}) when $\mathsf{p}_1$ is of the form (\ref{eq:2}).
Concerning the variance of the chopping time, it is given by
\begin{equation}
\label{eq:dtcdef}
(\delta t_c)^2:=A^2\left\langle (n-\langle n\rangle)^2\right\rangle=
A^2\sum_{n\geq1}(n-\langle n\rangle)^2\ \pi_n\ .
\end{equation}
Here again, an elementary computation based on the properties of the
geometric distribution (\ref{eq:pin}), yields $\delta
t_c=A\sqrt{1-P_1}/P_1$, so that the standard deviation of the chopping
time turns out to be given by
\begin{equation}
\label{eq:rhodef}
\delta t_c=A\ \frac{e^{N\ln\sqrt{1-\mathsf{p}_1}}}{
1-e^{N\ln(1-\mathsf{p}_1)}}\ ,
\end{equation}
of the form (\ref{eq:stdev}) when $\mathsf{p}_1$ is of the form
(\ref{eq:2}).  Notice that in the limit $N\to\infty$, $\mathsf{p}_1$
being kept fixed, $\delta t_c\to0$; on the other hand, if
$N\mathsf{p}_1\ll1$, one gets $\delta t_c\sim t_c$: the chain
chopping is a rare event and the fluctuations of the chopping time are
comparable with its mean.

\begin{figure}[h]
\includegraphics[height=8cm,width=12cm]{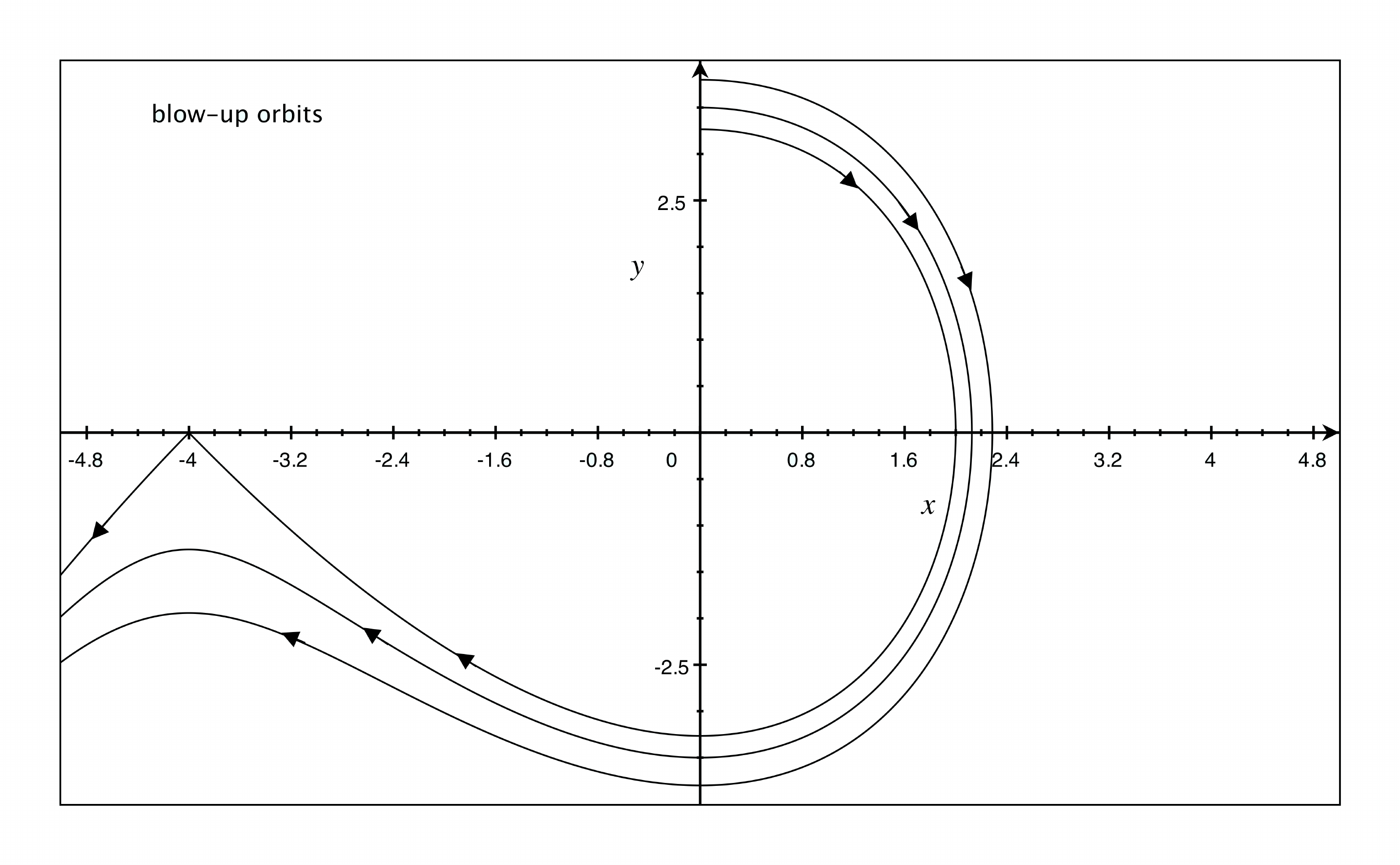}
\caption{\label{fig:3} Three phase curves of system (\ref{eq:rsddot})
  corresponding to the initial condition $r(0)=0$ and
  $\dot{r}(0)=+2\sqrt{\varepsilon}$.  The most internal one
  corresponds to the homoclinic connection value
  $\varepsilon=\DE=1/(6\alpha^2)=2.\bar6$.  The other two are blow-up
  curves, corresponding to $\varepsilon=3.06$ and $\varepsilon=3.61$.}
\end{figure}

We finally pass to the computation of $\mathsf{p}_1$, the probability,
according to the measure on the initial data, that a local blow-up
takes place in the chain on the time interval $[0,A]$, in order to
show that $\mathsf{p}_1$ is given by the expression (\ref{eq:2}) to a
rather good approximation. Obviously, \emph{$\mathsf{p}_1$ is the
  measure of the initial data such that the local blow-up time is less
  than $A$}.  In order to define the local blow-up time, let us
consider the equations of motion of the FPU $\alpha$-chain, with the
pair potential $V(r)=r^2/2+\alpha r^3/3$. The evolution equations of
the chain in terms of the variables $r_n:=q_{n+1}-q_n$ read
\begin{equation}
\label{eq:rddot}
\ddot{r}_n=V'(r_{n+1})+V'(r_{n-1})-2V'(r_n)\ ,
\end{equation}
whose form is valid for any potential $V$; as remarked above, we do
not discuss boundary effects. Let us consider a specific pair of sites
in the bulk, corresponding to $n=s$, where a blow-up event takes
place, namely $r_s\to-\infty$ in a finite time. In such a case, under
the hypothesis $|r_s|\gg|r_{s\pm1}|$, the main contribution to the
force on the right hand side of equation (\ref{eq:rddot}) is provided
by the last term, and one can consider the isolated two body problem
ruled by the equation
\begin{equation}
\label{eq:rsddot}
\ddot{r}_s=-2V'(r_s)\ .
\end{equation}
The latter equation, describing the relative dynamics of two nearby
particles, is the Newton equation of a single particle of mass $1/2$,
i.e. the reduced mass of two particles of unit mass, subject to the
force $-V'(r)$. The measure on the initial conditions on the particles
of the $\alpha$-chain is
\begin{equation}
\label{eq:dmu}
d\mu_0:=\left(\prod_{n=1}^N\sqrt{\frac{\beta}{\pi}} e^{-\beta p_n^2}
\delta(r_n)\right)dr_1\dots dr_N\ dp_1\dots dp_N\ ,
\end{equation}
where $p_n=\dot{r}_n/2$ is the relative momentum of two nearby
particles and $\delta(x)$ is the Dirac delta-function.  The measure
(\ref{eq:dmu}) corresponds to place all the particles in their
equilibrium position with a relative momentum distributed according to
the Maxwell-Boltzmann measure at temperature $T=\beta^{-1}$.  Now, in
the present specific case $V(r)=r^2/2+\alpha r^3/3$, and exploiting
the conservation of energy (at level $\varepsilon$), one easily finds
that the blow-up time $t_b$ of equation (\ref{eq:rsddot}), with
initial condition $r_s(0)=0$,
$p(0):=\dot{r}_s(0)/2=\pm\sqrt{\varepsilon}$, under the condition
$\varepsilon>\DE:=V(-1/\alpha)=1/(6\alpha^2)$, is given by the formula
\begin{equation}
\label{eq:tb}
t_{b}(\varepsilon)=\int_{-\infty}^0\frac{dr}{2\sqrt{\varepsilon-V(r)}}+
\theta(p(0))\int_0^{r_\varepsilon}\frac{dr}{\sqrt{\varepsilon-V(r)}}\ ,
\end{equation}
where $\theta(x)$ is the Heaviside step function, whereas
$r_\varepsilon$ is the value of the right turning point coordinate,
i.e. the (only) root of the equation $V(r)=\varepsilon$. The condition
$\varepsilon>\DE$ is obvious: the blow-up occurs along the energy
level curves that lie outside the homoclinic connection.  Notice that
the contribution of the second integral to the right hand side of
formula (\ref{eq:tb}) exists only when $p(0)=+\sqrt{\varepsilon}$,
because in this case the ``particle'' starts to move to the right,
reaches the turning point and then reverses the direction of motion
and goes to $-\infty$; see Figure~\ref{fig:3}.

Now we determine $\mathsf{p}_1(A)$ defined as the probability that
$t_b(\varepsilon)<A$ according to the measure (\ref{eq:dmu}).  Let us
observe that the blow-up time $t_b$, given by the formula
(\ref{eq:tb}), turns out to be a decreasing function of
$\varepsilon$. This means that the condition $t_b(\varepsilon)<A$ is
equivalent to $\varepsilon>\bar\varepsilon(A)$; moreover
$\bar\varepsilon\to\DE^+$ when $A\to+\infty$. Recalling that the
energy conservation law for problem (\ref{eq:rsddot}) reads
$p_s^2+V(r_s)=\varepsilon$, making use of the initial measure
(\ref{eq:dmu}) at the site of $n=s$, and omitting the site subscript
$s$, one finds
\begin{eqnarray}
\mathsf{p}_1(A)&=&\int_{\{p^2+V(r)>\bar\varepsilon(A)\}}
\sqrt{\frac{\beta}{\pi}}e^{-\beta p^2}
\delta(r)drdp=\nonumber\\ &=&\int_{\{p^2>\bar\varepsilon(A)\}}
\sqrt{\frac{\beta}{\pi}}\ e^{-\beta p^2}dp
=\frac{2}{\sqrt{\pi}}\int_{\sqrt{\beta\bar\varepsilon(A)}}^{+\infty}
e^{-y^2}dy=\nonumber\\ &=&\erfc\left(\sqrt{\beta\bar\varepsilon(A)}\right)\ ,
\label{eq:p1A}
\end{eqnarray}
where $\erfc(x):=(2/\sqrt{\pi})\int_x^{+\infty}e^{-y^2}dy$ is the
complementary error function.
The relation (\ref{eq:p1A}) depends clearly on $A$. However, on the basis of the previous remark, the blow-up time (\ref{eq:tb}) diverges on the
homoclinic connection, which means that
\begin{equation}
\label{eq:tbest}
t_b(\varepsilon)\sim f(\varepsilon-\DE)
\end{equation}
as $\varepsilon\to\DE^+$, where $f(x)$ is a monotonically decreasing
function such that $f(x)\to+\infty$ as $x\to0^+$.  As a consequence,
if the time step $A$ is large enough, the asymptotic behavior
(\ref{eq:tbest}) yields an estimate for $\bar\varepsilon(A)$, namely
\begin{equation}
\label{eq:barepsest}
\bar\varepsilon(A)\sim \DE+f^{-1}(A)\ ,
\end{equation}
for large values of $A$. In the latter approximation, the deviation of
the estimate (\ref{eq:barepsest}) with respect to $\DE$ is small,
since $f^{-1}(A)\to0^+$ as $A\to+\infty$. In conclusion, one can
reasonably choose $\bar\varepsilon(A)\simeq\DE$ in (\ref{eq:p1A}). As
a consequence,
\begin{equation}
\mathsf{p}_1(A)\simeq\mathsf{p}:=\erfc\left(\sqrt{\beta\DE}\right)\ ,
\end{equation}
which motivates the formula (\ref{eq:2}). Concerning the asymptotic
expansion of $\mathsf{p}$ when $\beta\DE$ is large, this is readily
obtained by the change of variable $y^2=\varepsilon$, which yields
\begin{eqnarray}
\label{pasymp}
\mathsf{p}&=&\frac{2}{\sqrt{\pi}}\int_{\sqrt{\beta\DE}}^{+\infty}
e^{-y^2}dy=\frac{1}{\sqrt{\pi}}\int_{\beta\DE}^{+\infty}
\frac{e^{-y}}{\sqrt{y}}dy=\nonumber\\ &=&\frac{1}{\sqrt{\pi}}\left[\frac{1}{(\beta\DE)^{1/2}}-
  \frac{1}{2(\beta\DE)^{3/2}}+O((\beta\DE)^{-5/2})\right]e^{-\beta\DE}\ ,
\end{eqnarray}
which explains formula (\ref{eq:3}). The approximation of large $A$
(whose actual numerical value is 4) is checked to hold a fortiori, by
the good agreement of the theoretical formulas with the numerical
data.

As already stressed at the end of the preceding Section, we find that the value of $\DE$ best fitting the
numerical data in Figure~\ref{fig:1} is $\DE=2.57$, which is a bit
lower than the theoretical one, namely $2.66$. In addition to what
remarked there, we here observe that the probability density of the $r_n$'s
is a delta function just at time $t=0$. At later times, the average
density will be some function $g(r)$ localized about $r=0$. As an
example, just to get an idea of the behavior of the above integral,
one can consider the simple case of a constant density
$g(r)=1/(2\eta)$ inside the interval $[-\eta,\eta]$, and $g(r)=0$
outside it, with $0<\eta\ll1$. In this case, the formula
(\ref{eq:p1A}), in the limit $A\to+\infty$, becomes
\begin{eqnarray}
\mathsf{p}(\beta)&=&\int_{\{p^2+V(r)>\DE\}}
\sqrt{\frac{\beta}{\pi}}e^{-\beta p^2} g(r)drdp=\nonumber\\
&=&\sqrt{\frac{\beta}{\pi}}\int_{\DE}^{+\infty}
e^{-\beta\varepsilon} \left(\int_{-\infty}^{+\infty}e^{\beta V(r)}
\frac{g(r)}{\sqrt{\varepsilon-V(r)}}\right)d\varepsilon\simeq\nonumber
\\
&\simeq&\sqrt{\frac{\beta}{\pi}}\int_{\DE}^{+\infty}  
\frac{e^{-\beta(\varepsilon-\eta^2/2)}}{\sqrt{\varepsilon-\eta^2/2}}
\ d\varepsilon = \erfc(\beta(\DE-\eta^2/2))\ .\nonumber
\end{eqnarray}
Thus, to the oscillation of the particles around their (initial)
equilibrium position, there corresponds a lower value of the effective
height of the energy barrier, as physical intuition would suggest.

\section{Final comments}\label{sec:4}

In the sequel, a four issues related to the subject of the present work are discussed.
 
Concerning the explanation of the lack of observed chain chopping of
the FPU $\alpha$-model in some numerical simulations, we stress that
in the present work we started from a crystal configuration, which
leads to the formula (\ref{eq:3}) at low temperature. For other
initial conditions, one has to take into account both the value of the
specific energy $\varepsilon$ and of the specific form of the initial
excitation. As an example, let us consider the one used by FPU in
their original paper \cite{FPU}, namely the initial excitation of the
lowest frequency mode with amplitude $B$ and the beaded string at
rest: $q_n^0=B\sin(\pi n/(N+1))$, $p_n^0=0$, $n=1,\dots,N$.  In such a
case, elementary calculations show that $T\simeq\varepsilon\simeq
\pi^2B^2/(4N^2)$, so that $\beta\DE\gg\ln N$ if $B\ll
2N\sqrt{\DE}/(\pi\sqrt{\ln N})$. On the other hand, one here has
$\sup|r_n|\simeq \pi B/N$, which is much lower than the critical value
$1/\alpha$ for the local blow-up if $B\ll N/(\pi\alpha)$. If $N$ is
large enough, the former condition implies the latter and the chain
will first reach the equipartition state, where
$r_n\approx\sqrt{\varepsilon}$, and then will be chopped on a very
long term (unless $\alpha^2\varepsilon\simeq 1$).  Just to give an
example, in the paper \cite{PCSF} the $\alpha$-model with $N=32$,
$\alpha=1/3$ and $E=1$, initialized with the above mentioned initial
excitation, was observed to reach and persist in the equipartition
state up to times of the order $10^8$. In this case, our formula
(\ref{eq:3}) would predict a chopping time $t_c\approx 10^{20}$, which
explains why the chain chopping was never observed in those runs.

\medskip

The estimate provided for the chopping time, could also be useful in
the case of glasses. In the glassy state, one usually assumes that the
microscopic dynamics remains trapped in a region of phase space,
corresponding to the glass phase, without entering the ``Boltzmann
sea'' corresponding to the crystal and/or the fluid phase. In fact particles are thought of as
frozen in the local minima of the potential, being able to jump out
only after a huge amount of time. A simple one--dimensional model in
which this happens is described in paper \cite{amati}. We are hopefully 
thinking to adapt the present analytic treatment to the study of
glasses.

\medskip

It has to be stressed that, contrary to what might appear at a first glance, the
condition $N\mathsf{p}\ll1$ for the validity of the asymptotic formula
(\ref{eq:3}), is completely meaningful from a physical point of
view. Indeed, another way of stating it is $T\ll\Delta E/\ln N$. The
latter, if the number of particles in the chain is $N=10^\gamma$,
reads $T\ll \Delta E/(\gamma\ln 10)$, the upper bound ranging from
$\Delta E/53$ for ordinary matter ($\gamma=23$) up to $\Delta E/14$
for a human gene ($\gamma=6$, the limit numerically explored in the
present paper). Thus, in dealing with macroscopic chains of a fixed
size, there always exists an observable temperature below which the
chain chopping, describing physical phenomena such as melting,
denaturation and so on, takes place on extremely long time-scales, on
average.

\medskip

Finally, we shortly discuss the possibility to approach the chain chopping problem with the theory developed by Kramers \cite{Kram}. A possible line of reasoning is the following.
The pair of particles where the chain breaks down is not an isolated system, being subject to the influence of the rest of
the chain. According to the picture emerging from the work of Ford, Kac and Mazur \cite{FKM}, one may suppose that the rest of the chain acts on
that pair of particles as an effective thermal bath, giving rise to
a Langevin dynamics of their relative displacement. In such a case one would deal with what is known in the literature as a \emph{Kramers problem}, namely the estimate of the escape time of a Brownian particle from a
potential well (see \cite{Mel} for a review on the
subject). The rate $w$ of escape over the potential
barrier, first computed by Kramers \cite{Kram}, turns out to be
given by an Arrhenius-like formula: $w\propto e^{-\beta\Delta E}$,
where $\beta=1/T$ is the inverse temperature of the thermal bath (here
proportional to the mean kinetic energy of the chain) and $\Delta E$
is the height of the potential barrier. On the other hand,
in a chain of $N$ sites, the rate of chain chopping would be
$Nw\propto Ne^{-\beta\Delta E}$, whose inverse gives the chopping
time, namely formula (\ref{eq:3}) up to a small correction (at small
temperature). However, in so reasoning, one meets two main difficulties. 
First of all, the $\alpha$-chain potential (\ref{eq:V}) is unbounded from below, while the Kramers
theory applies to stable potentials, those for which the Gibbs measure exists. Secondly, in the quoted work \cite{FKM}, 
a linear chain with a rather special frequency spectrum is considered, and an extension of those results to the
linearized FPU chain is not known to us.  In conclusion, a Kramers-like solution of the present problem may have only a heuristic character.

\centerline{\ldots\ldots}

\thanks{The authors thank L. Galgani for his useful comments on the manuscript.}

\end{document}